\documentclass{article}

\usepackage{PRIMEarxiv}
\usepackage{amsmath}
\usepackage[utf8]{inputenc} 
\usepackage[T1]{fontenc}    
\usepackage{hyperref}       
\usepackage{url}            
\usepackage{booktabs}       
\usepackage{amsfonts}       
\usepackage{nicefrac}       
\usepackage{microtype}      
\usepackage{lipsum}
\usepackage{fancyhdr}       
\usepackage{graphicx}       

\usepackage[table]{xcolor}

\graphicspath{{media/}}     

\title{Sensitivity and Early Detection of Bayesian Causal Impact Models for Marketing Interventions
}

\author{
  Jorge Pellegrini \\
  Marketing Science, Despegar \\
  Buenos Aires, Argentina \\  
  \texttt{jpellegrini@fra.utn.edu.ar} \\
  ORCID: \href{https://orcid.org/0000-0003-0526-7743}{0000-0003-0526-7743}
}

\begin{document}
\maketitle

\begin{abstract}
Marketing systems frequently undergo operational changes that may affect performance, making timely detection of adverse effects essential for decision making. While Bayesian Causal Impact models are widely used to estimate the causal effects of interventions, their ability to support early operational monitoring remains less explored. This paper proposes a simulation based framework to evaluate the sensitivity and detection capabilities of Bayesian causal impact analysis under controlled performance degradations. Using daily traffic data from an abandoned cart marketing journey, we repeatedly perturb the observed outcomes and assess alarm activation probabilities across different effect magnitudes, confidence levels, and detection horizons. Two alarm criteria are analyzed: one based on the proportion of observations falling below the predictive lower bound and another based on consecutive days of negative cumulative impact. Results show that detection performance depends strongly on the interaction between effect size, confidence level, and evaluation horizon. In particular, proportion based criteria become less effective as the monitoring horizon increases, whereas persistence based criteria provide more stable and operationally meaningful detection behavior. The proposed framework extends causal impact analysis beyond retrospective effect estimation, offering a practical methodology for quantifying detection sensitivity and supporting monitoring decisions in dynamic marketing environments.
\end{abstract}


\section{Introduction}
Digital marketing in e-commerce environments is inherently dynamic. Marketing teams continuously implement changes to strategies, systems, and automated workflows in response to evolving user behavior, competitive pressure, and business objectives\cite{mantha2020real,verhoef2021digital}. These changes may involve adjustments in the targeting rules, messaging logic, channel orchestration, or the underlying systems that govern how and when users are reached. As a consequence, marketing performance is shaped by an ongoing sequence of interventions rather than isolated static decisions\cite{koch2023dynamic,lemon2016understanding}.

Automated marketing journeys play a central role in e-commerce strategies. These journeys consist of predefined sequences of automated communications that are triggered when a user performs a specific action during their exploration process, as they move through the purchase funnel\cite{raina2023need}. One of the most prevalent and business-critical examples is the abandoned cart journey, which is activated when a user reaches the final stage of the funnel but does not complete the purchase. In such cases, a sequence of automated messages is triggered with the objective of re-engaging the user and recovering the conversion opportunity\cite{goic2021effectiveness}.

In practice, these automated journeys are constantly evolving. Marketing and engineering teams frequently introduce structural changes, such as replacing or upgrading the underlying software, deploying new system versions, or modifying the core execution logic. Unlike isolated campaign adjustments, these interventions operate at the system level and simultaneously affect the entire user population interacting with the journey. Consequently, it is typically not feasible to evaluate their impact through classical A/B testing, since controlled experimentation becomes impractical or impossible\cite{kohavi2020trustworthy}. A fundamental challenge emerges when trying to measure the true impact of a given structural change. Variations in key business metrics can reflect the effect of the intervention implemented , but they may also be driven by external factors not related to the change itself. Changes in demand, seasonality, market-wide trends, or changes in user intent can independently produce deterioration or improvement in performance metrics, complicating causal interpretation\cite{taiwo2024advanced}.

This context motivates the use of causal analysis techniques, whose objective is to attribute observed changes in outcomes to a specific intervention and quantify the effect of an action while accounting for confounding factors. However, the challenge extends beyond estimating whether an effect exists. From a business perspective, decision-makers require timely and reliable answers to operational questions: how many days of post-change data are needed to reach a confident conclusion, and how early it is possible to detect whether post-change performance is worse than pre-change behavior. These questions are essential for enabling actions such as rollbacks, system adjustments, or continued deployment\cite{kohavi2012trustworthy}.

Although existing causal analysis approaches are effective at estimating causal effects retrospectively, they provide limited guidance on the sensitivity of the analysis and its ability to support early decision-oriented assessments\cite{moraffah2021causal}. In particular, there remains a gap between causal estimation and the operational needs of marketing teams, who must balance uncertainty, timeliness, and business risk in highly dynamic environments.

Such a gap motivates the objective of this paper: to propose an operational framework that connects causal impact analysis with sensitivity assessment and early detection criteria, enabling actionable and reliable business decisions in dynamic marketing environments.

\section{Materials and Methods}
\label{sec:Materials and Methods}

The proposed methodology is designed to assess the sensitivity and early detection capabilities of causal impact analyses in operational marketing settings. Rather than relying on a single model fit, the approach evaluates the stability of causal conclusions under data perturbations, varying confidence levels, and controlled performance degradations. The framework is structured as a repeated simulation procedure that quantifies the probability of triggering an alarm under different detection horizons and uncertainty assumptions.

\subsection{Causal Impact Framework and Temporal Modeling Strategy}

Causal Impact is a time-series causal inference model designed to estimate the effect of an intervention. The method constructs a counterfactual trajectory for the target time series by exploiting its historical relationship with a set of control covariates that are predictive of the outcome but unaffected by the intervention. The central assumption underlying the approach is that this relationship remains stable over time, allowing patterns learned in the pre-intervention period to be extrapolated to the post-intervention window. Under this assumption, systematic deviations between the observed series and the predicted counterfactual are interpreted as causal effects attributable to the intervention\cite{brodersen2015inferring}.

Is based on a Bayesian structural time series model\cite{scott2014predicting}. The observed target series is decomposed into latent components that capture the underlying trends, seasonal patterns, and regression effects driven by external covariates. Formally, the outcome is represented as the sum of a latent state process and a regression term that links the target series to a set of contemporaneous covariates. Inference is conducted within a Bayesian framework in which uncertainty over both latent states and model parameters is quantified through posterior distributions. The regression component introduces coefficients that measure the contribution of each covariate to the target series, while the state-space formulation enables flexible modeling of temporal dynamics, including smooth trends and time-varying seasonal effects. Seasonality is explicitly modeled through dedicated state components, allowing the framework to account for recurring temporal structures commonly observed in marketing time series. Under the assumption that both the regression relationship and the latent state dynamics remain stable over time, posterior samples inferred from the pre-intervention period are propagated forward to generate a predictive distribution for the post-intervention window. This distribution defines the counterfactual trajectory of the target series in the absence of intervention, against which observed outcomes are compared to infer causal impact.

The model is trained using an initial time window, referred to as the training period, during which both the target series and a set of correlated covariates are observed. This period is used to learn the relationship between the outcome of interest and the covariates. The validation period corresponds to the final segment of the time series, which is held out from model fitting and used to generate counterfactual predictions.

For the implementation of the causal inference framework, this study employs the \texttt{CausalImpact} library \cite{brodersen2015inferring}, which provides a practical implementation of Bayesian structural time series models for intervention analysis.

\subsection{Data Structure and Experimental Design}

We consider a multivariate time-series setting in which the outcome variable represents the daily number of automated messages sent within an abandoned cart journey, which is activated when a user reaches the final stage of the purchase funnel, but does not complete the transaction. The covariates correspond to business activity signals related to the same product category, including the daily number of users entering abandoned cart flows and the overall volume of product-related searches on the platform. These variables are expected to capture the underlying demand dynamics and user intent while remaining unaffected by the intervention itself.

The analysis is motivated by the need to evaluate the impact of a structural change in the automated journey delivery mechanism. In particular, the objective is not only to assess whether the change produces a measurable effect, but also to determine how many days of post-change data are required to obtain a reliable assessment of performance and to identify potential deterioration in a timely manner.

From an operational perspective, this translates into the need for an alarm system capable of signaling whether post-change performance deviates negatively from pre-change behavior. Within the validation window, the observed outcome is artificially reduced by a given percentage, representing a simulated loss of efficiency after the structural change. This design enables the evaluation of detection mechanisms under controlled effect sizes and allows us to study how quickly and reliably performance deterioration can be identified.

\subsection{Subsampling and Data Perturbation}

To capture the sensitivity of the model to data variability and structural uncertainty, we adopt a repeated subsampling and perturbation strategy\cite{efron1994introduction}. For each iteration, the original time series is rescaled using a subsampling factor and augmented with stochastic noise proportional to the magnitude of each series. This procedure preserves the temporal structure while introducing realistic fluctuations in both the outcome and the covariates.

This step is repeated across multiple iterations, generating a collection of perturbed datasets that reflect plausible alternative realizations of the observed data. 

\subsection{Repeated Causal Impact Estimation and Synthetic Impact Injection }

For each perturbed dataset, a causal impact analysis is performed using a fixed pre-period and post-period definition aligned with the training and validation windows. The model produces a counterfactual prediction of the outcome during the validation period, along with posterior uncertainty summarized through point estimates and credible intervals.

Rather than relying exclusively on the default 95 credible interval, we reconstruct predictive intervals at multiple confidence levels by leveraging the implied posterior distribution. This allows the evaluation of detection behavior under different degrees of statistical conservatism.

To simulate adverse post-change scenarios, we introduce controlled synthetic reductions in the observed outcome during the validation period. These reductions are defined as multiplicative factors applied to the observed series, representing varying magnitudes of performance degradation. This procedure enables the evaluation of detection performance across a range of effect sizes without altering the underlying covariate dynamics.

\subsection{Alarm Definition and Detection Criteria}
For each iteration, confidence level, reduction magnitude, and detection horizon, we define an alarm criterion based on the proportion of days in which the observed (reduced) outcome falls below the lower bound of the predicted credible interval. The detection horizon is defined as the number of elapsed days after the intervention available for evaluation. An alarm is triggered if this proportion exceeds a predefined threshold within the specified horizon. Similar threshold-based monitoring approaches are widely used in statistical process control and anomaly detection systems to identify persistent deviations from expected behavior\cite{montgomery2020introduction,chandola2009anomaly}.

This rule captures both the persistence and magnitude of deviations, reflecting realistic operational decision criteria rather than isolated pointwise violations.

Across all simulation iterations, we compute the probability of triggering an alarm for each combination of reduction magnitude, confidence level, and detection horizon. These probabilities quantify the sensitivity of the causal impact analysis to adverse performance changes and provide an interpretable measure of detection power under operational constraints.

\subsection{Alarm Design}

Let $y_t$ denote the observed outcome at time $t$, representing the number of automated journey sends, and let
\[
\mathbf{x}_t = (x_{1t}, \dots, x_{pt})
\]
denote a vector of covariates capturing business activity related to abandoned carts for the same product. The observed data consist of a multivariate time series
\[
\{(y_t, \mathbf{x}_t)\}_{t=1}^T .
\]

The observation window is partitioned into a training period
\[
\mathcal{T}_{\text{train}} = \{1, \dots, T_0\}
\]
and a validation period
\[
\mathcal{T}_{\text{val}} = \{T_0+1, \dots, T\}.
\]

To assess sensitivity to data variability, we generate perturbed versions of the observed series. For each iteration $k = 1, \dots, N$, we construct
\[
y_t^{(k)} = \alpha \, y_t + \varepsilon_t^{(k)}, \quad
x_{jt}^{(k)} = \alpha \, x_{jt} + \eta_{jt}^{(k)},
\]
where $\alpha \in (0,1)$ is a subsampling factor. In the experiments presented in this study, the subsampling factor was fixed at $\alpha = 0.7$. The perturbation terms
\[
\varepsilon_t^{(k)} \sim \mathcal{N}(0, \sigma^2 y_t^2), \quad
\eta_{jt}^{(k)} \sim \mathcal{N}(0, \sigma^2 x_{jt}^2)
\]
introduce relative stochastic perturbations, with the relative noise level set to $\sigma = 0.05$ in all simulations. This construction preserves the temporal structure of the data while reflecting realistic, scale-dependent operational noise.

For each perturbed dataset $\{y_t^{(k)}, \mathbf{x}_t^{(k)}\}$, a Bayesian structural time series model is fitted on $\mathcal{T}_{\text{train}}$. The model yields a posterior predictive distribution for the counterfactual outcome during the validation period:
\[
y_t^{(k)} \mid \mathbf{x}_t^{(k)} \sim
\mathcal{N}\!\left(\hat{\mu}_t^{(k)}, \hat{\sigma}_t^{2(k)}\right),
\quad t \in \mathcal{T}_{\text{val}}.
\]

For a given confidence level $c \in (0,1)$, the lower predictive bound is defined as
\[
L_{t,c}^{(k)} =
\hat{\mu}_t^{(k)} + z_{(1-c)/2} \, \hat{\sigma}_t^{(k)},
\]
where $z_{(1-c)/2}$ denotes the corresponding Gaussian quantile.

To simulate post-change performance deterioration, we define a degraded observed series
\[
\tilde{y}_t^{(k,r)} = r \, y_t^{(k)}, \quad r \in (0,1],
\]
where $r$ represents the magnitude of efficiency loss. This construction allows evaluation of detection behavior under controlled effect sizes.

Let $h$ denote a detection horizon (number of days). For each combination $(k, r, c, h)$, we compute
\[
\phi_{k,r,c,h}
=
\frac{1}{h}
\sum_{t=T_0+1}^{T_0+h}
\mathbb{I}\!\left(
\tilde{y}_t^{(k,r)} < L_{t,c}^{(k)}
\right),
\]
which represents the fraction of days within the horizon for which the degraded outcome falls below the lower predictive bound.

An alarm indicator is then defined as
\[
A_{k,r,c,h}
=
\mathbb{I}\!\left(
\phi_{k,r,c,h} > \tau
\right),
\]
where $\tau \in (0,1)$ is a predefined threshold representing the minimum proportion of violations required to trigger an alarm.

The causal impact model provides predictive uncertainty in the form of credible pointwise intervals at a fixed confidence level of $95\%$. Let $[\hat{L}_{t,95}^{(k)}, \hat{U}_{t,95}^{(k)}]$ denote the lower and upper bounds of this interval for iteration $k$ at time $t$. Under the Gaussian approximation implied by the posterior predictive distribution, the predictive variance can be recovered as
\[
\hat{\sigma}_t^{(k)} =
\frac{\hat{U}_{t,95}^{(k)} - \hat{L}_{t,95}^{(k)}}{2 z_{0.975}},
\]
where $z_{0.975}$ denotes the $97.5$th percentile of the standard normal distribution.

Using this estimated variance, predictive bounds for any confidence level $c \in (0,1)$ can be reconstructed as
\[
L_{t,c}^{(k)} = \hat{\mu}_t^{(k)} + z_{(1-c)/2} \, \hat{\sigma}_t^{(k)}, \quad
U_{t,c}^{(k)} = \hat{\mu}_t^{(k)} + z_{(1+c)/2} \, \hat{\sigma}_t^{(k)}.
\]

This procedure allows the evaluation of detection behavior under multiple confidence levels without refitting the model, enabling a systematic exploration of the trade-off between statistical conservatism and early detection sensitivity.

Finally, the detection probability is estimated as
\[
P_{r,c,h}
=
\frac{1}{N}
\sum_{k=1}^{N}
A_{k,r,c,h},
\]
which quantifies the likelihood of detecting a performance deterioration of magnitude $r$ within $h$ days at confidence level $c$.

The resulting surface $P_{r,c,h}$ characterizes the sensitivity of the causal impact analysis as a function of effect size, uncertainty tolerance, and detection latency. This formulation provides a principled mapping between statistical uncertainty and operational decision-making, directly addressing how much data is required to reliably detect a given level of performance degradation.

\section{Results}
The case study is based on six months of daily traffic data from Despegar's marketing platform. The analysis focuses on a specific user population associated with an abandoned cart journey, for which the outcome variable exhibits an average daily volume of approximately 300 automated messages. The first five months are used as a training period, while the last month is reserved as a validation window.

The full observation window spans six months of daily data, drawn from a population with an average daily message volume of 300. The first five months are used as a training period, while the last month is reserved as a validation window.

Sensitivity analysis is conducted by evaluating the probability of triggering an alarm in different configurations of performance degradation, confidence levels, and detection horizons.

Synthetic performance reductions are applied to the observed outcome during the validation period using multiplicative factors
\[
r \in \{1.0, 0.9, 0.8, 0.7, 0.5\},
\]
where $r=1.0$ represents the baseline scenario with no degradation, and lower values represent increasing levels of performance loss.

Predictive intervals are evaluated at multiple confidence levels
\[
c \in \{60\%, 70\%, 80\%, 90\%\},
\]
allowing analysis of the trade-off between statistical conservatism and detection sensitivity.

To assess how quickly a degradation can be detected, the validation period is truncated at different horizons
\[
h \in \{5, 7, 10, 15\} \text{ days},
\]
simulating the number of days of post-intervention data available for decision-making.

For each configuration, the alarm is based on the proportion of days within the evaluation horizon for which the observed value falls below the lower bound of the predictive interval.

Formally, let $\tilde{y}_t^{(r)}$ denote the degraded observed series and $L_{t,c}$ the lower bound of the predictive interval at confidence level $c$. The fraction of observations falling below the lower predictive bound
over a horizon $h$ is defined as

\[
\phi_{r,c,h} = \frac{1}{h} \sum_{t=1}^{h} \mathbb{I} \left( \tilde{y}_t^{(r)} < L_{t,c} \right).
\]

An alarm is triggered when this fraction exceeds a predefined threshold $\tau = 0.4$:
\[
A_{r,c,h} = \mathbb{I}(\phi_{r,c,h} > 0.4).
\]

An alarm is triggered when more than 40\% of the observations within the evaluation horizon fall below the lower predictive bound. For the shortest horizon considered (5 days), this corresponds to at least 3 observations.

Table~\ref{tab:sensitivity_results} reports the empirical probability of triggering an alarm across all configurations. For each configuration, the probability was estimated from 100 simulation runs.

\newcommand{\Prob}[1]{%
  \cellcolor{red!#1}#1\%
}

\begin{table}[ht]
\centering
\scriptsize
\caption{Probability of triggering an alarm ($P_{r,c,h}$) by reduction level, confidence interval, and detection horizon.}
\resizebox{\textwidth}{!}{
\begin{tabular}{c|cccc|cccc|cccc|cccc}
\hline
 & \multicolumn{4}{c}{5 days} & \multicolumn{4}{c}{7 days} & \multicolumn{4}{c}{10 days} & \multicolumn{4}{c}{15 days} \\
Reduction
& IC90 & IC80 & IC70 & IC60
& IC90 & IC80 & IC70 & IC60
& IC90 & IC80 & IC70 & IC60
& IC90 & IC80 & IC70 & IC60 \\
\hline

$\times 1.0$
& \Prob{0} & \Prob{5} & \Prob{30} & \Prob{65}
& \Prob{28} & \Prob{78} & \Prob{95} & \Prob{100}
& \Prob{0} & \Prob{0} & \Prob{0} & \Prob{2}
& \Prob{0} & \Prob{0} & \Prob{0} & \Prob{0} \\

$\times 0.9$
& \Prob{0} & \Prob{15} & \Prob{65} & \Prob{85}
& \Prob{60} & \Prob{88} & \Prob{100} & \Prob{100}
& \Prob{0} & \Prob{0} & \Prob{2} & \Prob{12}
& \Prob{0} & \Prob{0} & \Prob{0} & \Prob{2} \\

$\times 0.8$
& \Prob{2} & \Prob{55} & \Prob{82} & \Prob{95}
& \Prob{78} & \Prob{100} & \Prob{100} & \Prob{100}
& \Prob{0} & \Prob{2} & \Prob{10} & \Prob{45}
& \Prob{0} & \Prob{0} & \Prob{2} & \Prob{35} \\

$\times 0.7$
& \Prob{15} & \Prob{78} & \Prob{92} & \Prob{98}
& \Prob{88} & \Prob{100} & \Prob{100} & \Prob{100}
& \Prob{0} & \Prob{5} & \Prob{40} & \Prob{75}
& \Prob{0} & \Prob{0} & \Prob{38} & \Prob{90} \\

$\times 0.5$
& \Prob{78} & \Prob{95} & \Prob{100} & \Prob{100}
& \Prob{98} & \Prob{100} & \Prob{100} & \Prob{100}
& \Prob{8} & \Prob{62} & \Prob{95} & \Prob{100}
& \Prob{8} & \Prob{70} & \Prob{98} & \Prob{100} \\

\hline
\end{tabular}
}
\label{tab:sensitivity_results}
\end{table}

The results presented in Table~\ref{tab:sensitivity_results} do not exhibit a fully consistent detection pattern across all configurations. While, as expected, the probability of triggering an alarm increases with the magnitude of the reduction, the behavior across confidence levels and horizons reveals important limitations of the proposed criterion.

In particular, the requirement that more than 40\% of the days within the evaluation horizon fall below the lower predictive bound introduces a structural bias. As the horizon increases, the criterion becomes increasingly difficult to satisfy, since it requires a sustained proportion of observations below the lower predictive bound over a larger number of observations. As a result, for longer horizons (e.g., 10 or 15 days), the alarm is only triggered under very strong reductions or when using low confidence levels, which effectively narrows the predictive interval.

This behavior limits the practical applicability of the criterion in operational settings, where early detection under moderate degradations is critical.

To address this limitation, an alternative alarm definition is proposed based on temporal persistence rather than aggregate proportion. Specifically, an alarm is triggered if the cumulative impact remains negative for three consecutive days. Formally, defining the daily impact as
\[
\Delta_t = \tilde{y}_t^{(r)} - \hat{\mu}_t,
\]
the alarm condition is given by the existence of a sequence
\[
\Delta_t < 0, \quad \Delta_{t+1} < 0, \quad \Delta_{t+2} < 0.
\]

The results obtained under this alternative criterion are presented in Table~\ref{tab:sensitivity_consecutive}.

Compared to the previous approach, this criterion yields a more stable and operationally meaningful detection behavior. By focusing on consecutive deviations rather than overall proportions, it avoids dilution effects that arise in longer horizons and allows the detection mechanism to respond more quickly to sustained negative trends. Incorporating temporal persistence is a common strategy in monitoring and change-detection frameworks to reduce sensitivity to transient fluctuations and random noise\cite{basseville1993detection}.

Additionally, the results show that excessively low confidence levels lead to a rapid increase in alarm probability even under mild reductions. This is expected, as narrower predictive intervals increase the likelihood of observing violations. Therefore, while lower confidence levels improve sensitivity, they may also increase the risk of false positives.

From a business perspective, the results suggest that a balanced configuration can be achieved by using a detection horizon of 10 days combined with a confidence level of 80\%. This setup provides a reasonable trade-off between robustness and sensitivity, avoiding excessive false alarms while still enabling the detection of performance deteriorations that are relevant from an operational standpoint.

\begin{table}[ht]
\centering
\scriptsize
\caption{Probability of triggering an alarm based on three consecutive days of negative cumulative impact ($P_{r,c,h}$).}
\resizebox{\textwidth}{!}{
\begin{tabular}{c|cccc|cccc|cccc|cccc}
\hline
 & \multicolumn{4}{c}{5 days} & \multicolumn{4}{c}{7 days} & \multicolumn{4}{c}{10 days} & \multicolumn{4}{c}{15 days} \\
Reduction
& IC90 & IC80 & IC70 & IC60
& IC90 & IC80 & IC70 & IC60
& IC90 & IC80 & IC70 & IC60
& IC90 & IC80 & IC70 & IC60 \\
\hline

$\times 1.0$
& \Prob{0} & \Prob{0} & \Prob{0} & \Prob{0}
& \Prob{0} & \Prob{0} & \Prob{12} & \Prob{95}
& \Prob{0} & \Prob{0} & \Prob{12} & \Prob{95}
& \Prob{0} & \Prob{0} & \Prob{12} & \Prob{95} \\

$\times 0.9$
& \Prob{0} & \Prob{0} & \Prob{0} & \Prob{22}
& \Prob{0} & \Prob{5} & \Prob{82} & \Prob{100}
& \Prob{0} & \Prob{5} & \Prob{82} & \Prob{100}
& \Prob{0} & \Prob{5} & \Prob{82} & \Prob{100} \\

$\times 0.8$
& \Prob{0} & \Prob{0} & \Prob{18} & \Prob{78}
& \Prob{0} & \Prob{50} & \Prob{100} & \Prob{100}
& \Prob{0} & \Prob{50} & \Prob{100} & \Prob{100}
& \Prob{0} & \Prob{50} & \Prob{100} & \Prob{100} \\

$\times 0.7$
& \Prob{0} & \Prob{15} & \Prob{80} & \Prob{100}
& \Prob{2} & \Prob{95} & \Prob{100} & \Prob{100}
& \Prob{2} & \Prob{95} & \Prob{100} & \Prob{100}
& \Prob{2} & \Prob{95} & \Prob{100} & \Prob{100} \\

$\times 0.5$
& \Prob{20} & \Prob{100} & \Prob{100} & \Prob{100}
& \Prob{88} & \Prob{100} & \Prob{100} & \Prob{100}
& \Prob{88} & \Prob{100} & \Prob{100} & \Prob{100}
& \Prob{88} & \Prob{100} & \Prob{100} & \Prob{100} \\

\hline
\end{tabular}
}
\label{tab:sensitivity_consecutive}
\end{table}

\section{Conclusion}

This study proposes an operational framework to evaluate the sensitivity and early detection capabilities of Bayesian causal impact models in dynamic marketing environments. By combining repeated perturbation-based simulations, synthetic performance degradations, and alternative alarm criteria, the framework extends traditional causal impact analysis beyond retrospective effect estimation to actionable operational monitoring.

The results show that detection performance strongly depends on the interaction between effect magnitude, confidence level, and evaluation horizon. In particular, alarm definitions based on aggregate proportions of violations become less effective as the detection horizon increases, limiting their applicability for timely operational decisions. In contrast, persistence-based criteria that rely on consecutive negative deviations provide a more stable and practically meaningful detection behavior.

From a business perspective, the proposed methodology offers a principled mechanism for quantifying how quickly and reliably performance deteriorations can be detected following a structural intervention. This enables marketing and engineering teams to better balance sensitivity, robustness, and false-alarm risk when defining monitoring strategies for automated journeys and large-scale system changes.

More broadly, the framework establishes a bridge between Bayesian causal inference and operational decision-making, providing a practical approach to transforming predictive uncertainty into actionable monitoring policies within real-world marketing systems.

Future research should focus on evaluating the temporal stability of the proposed alarm criteria. While the present study assesses detection performance using a fixed validation window, operational deployment requires understanding whether alarm signals remain consistent under different temporal realizations of the data. A natural extension is to incorporate an additional dimension based on rolling or shifting validation windows, repeatedly applying the alarm criteria across multiple window positions, following principles commonly adopted in rolling-origin evaluation for
time-series analysis\cite{tashman2000out}.
This would enable the assessment of alarm consistency and robustness over time, providing a more reliable characterization of detection performance. Such an approach could help distinguish persistent deterioration signals from window-specific fluctuations, ultimately leading to more stable and trustworthy monitoring policies in real-world applications.

\bibliographystyle{unsrt}  
\bibliography{references}

@article{brodersen2015inferring,
  title={Inferring causal impact using Bayesian structural time-series models},
  author={Brodersen, Kay H and Gallusser, Fabian and Koehler, Jim and Remy, Nicolas and Scott, Steven L},
  year={2015}
}

@inproceedings{mantha2020real,
  title={A real-time whole page personalization framework for E-commerce},
  author={Mantha, Aditya and Sundaresan, Anirudha and Kedia, Shashank and Arora, Yokila and Gupta, Shubham and Wang, Gaoyang and Kanumala, Praveenkumar and Guo, Stephen and Achan, Kannan},
  booktitle={2020 IEEE International Conference on Big Data (Big Data)},
  pages={4646--4650},
  year={2020},
  organization={IEEE}
}

@article{koch2023dynamic,
  title={Dynamic customer journey analysis and its advertising impact},
  author={Koch, Christian and Lindenbeck, Benedikt and Olbrich, Rainer},
  journal={Journal of Strategic Marketing},
  pages={1--20},
  year={2023},
  publisher={Taylor \& Francis}
}

@inproceedings{raina2023need,
  title={The need for marketing automation: A review},
  author={Raina, Attideep and Lamkuche, Hemraj},
  booktitle={AIP Conference Proceedings},
  volume={2914},
  number={1},
  pages={030013},
  year={2023},
  organization={AIP Publishing LLC}
}

@article{goic2021effectiveness,
  title={The effectiveness of triggered email marketing in addressing browse abandonments},
  author={Goic, Marcel and Rojas, Andrea and Saavedra, Ignacio},
  journal={Journal of Interactive Marketing},
  volume={55},
  number={1},
  pages={118--145},
  year={2021},
  publisher={SAGE Publications Sage CA: Los Angeles, CA}
}

@book{kohavi2020trustworthy,
  title={Trustworthy online controlled experiments: A practical guide to a/b testing},
  author={Kohavi, Ron and Tang, Diane and Xu, Ya},
  year={2020},
  publisher={Cambridge University Press}
}

@article{taiwo2024advanced,
  title={Advanced A/B testing and causal inference for AI-driven digital platforms: A comprehensive framework for US digital markets},
  author={Taiwo, Kamorudeen Abiola and Akinbode, Azeez Kunle and Uchenna, E},
  journal={International Journal of Computer Applications Technology and Research},
  volume={13},
  number={6},
  pages={24--46},
  year={2024}
}

@inproceedings{kohavi2012trustworthy,
  title={Trustworthy online controlled experiments: Five puzzling outcomes explained},
  author={Kohavi, Ron and Deng, Alex and Frasca, Brian and Longbotham, Roger and Walker, Toby and Xu, Ya},
  booktitle={Proceedings of the 18th ACM SIGKDD international conference on Knowledge discovery and data mining},
  pages={786--794},
  year={2012}
}

@article{moraffah2021causal,
  title={Causal inference for time series analysis: Problems, methods and evaluation},
  author={Moraffah, Raha and Sheth, Paras and Karami, Mansooreh and Bhattacharya, Anchit and Wang, Qianru and Tahir, Anique and Raglin, Adrienne and Liu, Huan},
  journal={Knowledge and Information Systems},
  volume={63},
  number={12},
  pages={3041--3085},
  year={2021},
  publisher={Springer}
}

@article{scott2014predicting,
  title={Predicting the present with Bayesian structural time series},
  author={Scott, Steven L and Varian, Hal R},
  journal={International Journal of Mathematical Modelling and Numerical Optimisation},
  volume={5},
  number={1-2},
  pages={4--23},
  year={2014},
  publisher={Inderscience Publishers Ltd}
}

@book{efron1994introduction,
  title={An introduction to the bootstrap},
  author={Efron, Bradley and Tibshirani, Robert J},
  year={1994},
  publisher={Chapman and Hall/CRC}
}

@book{montgomery2020introduction,
  title={Introduction to statistical quality control},
  author={Montgomery, Douglas C},
  year={2020},
  publisher={John wiley \& sons}
}

@article{chandola2009anomaly,
  title={Anomaly detection: A survey},
  author={Chandola, Varun and Banerjee, Arindam and Kumar, Vipin},
  journal={ACM computing surveys (CSUR)},
  volume={41},
  number={3},
  pages={1--58},
  year={2009},
  publisher={ACM New York, NY, USA}
}

@book{basseville1993detection,
  title={Detection of abrupt changes: theory and application},
  author={Basseville, Michele and Nikiforov, Igor V and others},
  volume={104},
  year={1993},
  publisher={Prentice hall Englewood Cliffs}
}

@article{tashman2000out,
  title={Out-of-sample tests of forecasting accuracy: an analysis and review},
  author={Tashman, Leonard J},
  journal={International journal of forecasting},
  volume={16},
  number={4},
  pages={437--450},
  year={2000},
  publisher={Elsevier}
}

@article{verhoef2021digital,
  title={Digital transformation: A multidisciplinary reflection and research agenda},
  author={Verhoef, Peter C and Broekhuizen, Thijs and Bart, Yakov and Bhattacharya, Abhi and Dong, John Qi and Fabian, Nicolai and Haenlein, Michael},
  journal={Journal of business research},
  volume={122},
  pages={889--901},
  year={2021},
  publisher={Elsevier}
}

@article{lemon2016understanding,
  title={Understanding customer experience throughout the customer journey},
  author={Lemon, Katherine N and Verhoef, Peter C},
  journal={Journal of marketing},
  volume={80},
  number={6},
  pages={69--96},
  year={2016},
  publisher={SAGE Publications Sage CA: Los Angeles, CA}
}

\end{document}